\documentstyle[12pt,aaspp4]{article}
\begin{document}

\title{WHERE IS THE CORONAL LINE REGION IN ACTIVE GALACTIC NUCLEI ?}

\author{\sc Takashi Murayama 
        and Yoshiaki Taniguchi}
\affil{Astronomical Institute, Tohoku University,
                Aoba, Sendai 980-8578, Japan;
       murayama@astr.tohoku.ac.jp, tani@astr.tohoku.ac.jp}

\begin{abstract}
We report the new finding that type 1 Seyfert nuclei (S1s) have 
excess [FeVII]$\lambda$6087 emission with respect to type 2s (S2s).
The S1s exhibit broad emission lines which are attributed to ionized gas within
1 pc of the black hole, whereas the S2s do not show such broad lines.
The current unified model of active galactic nuclei explains this difference as that 
the central 1 pc region in the S2s is hidden 
from the line of sight by a dusty torus 
if we observe it from a nearly edge-on view toward the torus.
Therefore, our finding implies that the coronal line region (CLR)
traced by the [FeVII]$\lambda$6087 emission  resides in the inner wall of
such dusty tori.
On the other hand, the frequency of occurrence of the CLR in the optical
spectra is nearly the same between the S1s and the S2s.
Moreover, some Seyfert nuclei exhibit a very extended ($\sim$ 1 kpc) CLR.
All these observational results can be unified if we introduce a three-component
model for the CLR;
1) the inner wall of the dusty torus,
2) the clumpy ionized region associated with the narrow line region
at distance from $\sim$ 10 to $\sim$ 100 pc, and 3) the extended
ionized region at distance $\sim$ 1 kpc. 

\end{abstract}



\section{INTRODUCTION}

Optical spectra of active galactic nuclei (AGN) show often very high ionization
emission lines such as [FeVII], [FeX], and [FeXVI] (the so-called coronal lines).
Since the ionization potentials of these lines are
higher than 100 eV, much attention has been paid to
the understanding of the coronal line region (CLR)
(e.g., Oke \& Sargent 1968; Osterbrock 1977, 1985; Grandi 1978; Pelat et al. 1981; 
Penston et al. 1984).
It is often considered that the CLR has an intermediate nature 
between the broad line region (BLR) and the 
narrow line region (NLR) because the high-ionization lines
have critical densities for collisional excitation  are of the order of $10^7$ cm$^{-3}$
and some Seyfert nuclei show CLR emission lines with FWHM $\sim$ 1000 - 2000 km s$^{-1}$
(De Robertis \& Osterbrock 1984, 1986; Veilleux 1988; Appenzellar \& \"Osreicher 1988;
Appenzellar \& Wagner 1991).

Recent photoionization model calculations have suggested that 
the CLR is located mostly  within inner 10 pc (Ferguson, Korista, \& Ferland 1997).
In fact, Oliva et al. (1994) found a compact ($<$ 10 pc) CLR 
in nearby Seyfert, the Circinus galaxy, using the near infrared
coronal line [SiVI] at 1.92 $\mu$m.
However, it is also known that some Seyfert nuclei have an extended CLR
whose size amounts up to $\sim$ 1 kpc [Golev et al. 1994; Murayama, Taniguchi,
\& Iwasawa 1998 (hereafter MTI98)].
The presence of such extended CLRs is usually explained 
as the result of very low-density conditions in the interstellar medium 
($n_{\rm H} \sim 1$ cm$^{-3}$)
makes it possible to achieve higher ionization conditions
(Korista \& Ferland 1989).
The above complicated situation raises the  question;
{\it Where is the CLR in AGN ?}

According to the current unified models(Antonucci \& Miller 1985; Antonucci 1993), 
it is generally believed that
a dusty torus surrounds both the central engine and the BLR.
Since the inner wall of the torus is exposed to intense radiation
from the central engine, it is naturally expected that the wall 
can be one of the important sites for the CLR (Pier \& Voit 1995).
Recently, Gallimore et al. (1997) discovered a very compact ($\sim$ 1 pc) 
ionized region in the S2 galaxy NGC 1068 in radio continuum.
Since the inner radius of the accreting molecular ring traced by water vapor maser
emission  is $\sim$ 0.5 pc (Greenhill et al. 1996), 
this ionized region seems indeed to be the
inner wall of the torus.
If the inner wall is an important site of CLR, 
it should be  expected that the S1s would tend to have 
more intense CLR emission because the inner wall would be  obscured by the torus itself
in S2s.
In order to examine whether or not the S1s tend to have the excess CLR
emission, we study the frequency distributions of the
[FeVII]$\lambda$6087/[OIII]$\lambda$5007 intensity ratio between S1s and S2s.

\section{DATA AND RESULTS}

The data were compiled from the literature (Osterbrock 1977, 1985; Koski 1978;
Osterbrock \& Pogge 1985; Shuder \& Osterbrock 1981)
and our own optical spectroscopic  data
of one S1 (NGC 4051) and four S2s (NGC 591, NGC 5695, NGC 5929, and NGC 5033)
which were taken with a CCD spectrograph attached to the Cassegrain
focus of the 188 cm telescope at Okayama Astrophysical Observatory.
In total, our sample contains 18 S1s and 17 S2s.
Although the sample is not a statistically complete one in any sense,
the data set is the largest one for the study of CLR ever compiled.
The average redshifts are similar between the S1s (0.031$\pm$0.017)
and  the S2s (0.024$\pm$0.016). There is no correlation between the redshift and 
the [FeVII]/[OIII] intensity ratio for both samples.

The result is shown in Figure 1.
It is shown that the S1s are strong [FeVII] emitters than the S2s.
Adopting the null hypothesis that the S1s and S2s studied here come from the same
underlying population, the Kormogrov-Smirnov statistical
 test gives that the probability of randomly 
selecting the observed ratios from the same population is only 5.0$\times 10^{-7}$.
Therefore the excess [FeVII] emission in the S1s with respect to the S2s
is statistically real.
In order to verify that this difference is really due to the excess [FeVII]
emission, we compare the [OIII] luminosity between the S1s and S2s and find that 
the [OIII] luminosity distribution is nearly the same between 
the S1s and the S2s (Figure 2).
Therefore, we conclude that the higher [FeVII]/[OIII] intensity ratio in the S1s 
is indeed due to the excess [FeVII] emission  rather than the weaker
[OIII] emission in the S1s.  
The presence of an excess [FeVII] emission in S1s can only be explained if
there is a fraction of the  inner CLR that cannot be seen in the S2s.
The height of the inner wall is of order 1 pc (Gallimore et al. 1997;
Pier \& Krolik 1992, 1993).
Therefore, given that the torus obscures this CLR from our line of sight,
the effective height of the torus should be significantly higher than 1 pc.

Although our new finding suggests strongly that part of the CLR emission arises
from the inner walls of dusty tori,
it is remembered that a number of S2s have the CLR.
In fact, the fraction of Seyfert nuclei with the CLR
is nearly the same between the S1s and the S2s (Osterbrock 1977; Koski 1978).
If the CLR was mostly concentrated in the inner 1 pc region, we would observe
the CLR only in the S1s. 
Therefore the presence of CLR in the S2s implies that there is another CLR 
component which has no viewing-angle dependence. 
A typical dimension of such a component is of order 100 pc like that of the NLR.
Ferguson et al. (1997) showed theoretically that 
the CLR can arise from just outside the 
BLR to $\sim 400 L_{43.5}^{1/2}$ pc where $L_{43.5}$ is the ionizing continuum
luminosity in units of 10$^{43.5}$ erg s$^{-1}$.
Since this size is almost comparable with that of the NLR, 
it is considered that a substantial part of the CLR coexists with the NLR.
MTI98 showed that the CLR of the high-ionization
Seyfert galaxy Tololo 0109$-$383 is spatially extended up to
a radius of 1.1 kpc. However, $\sim$ 70 \% of the CLR emission
is concentrated in the inner 220 pc in radius.
Since the ionizing continuum luminosity of Tololo 0109$-$383, inferred from the
bolometric luminosity, is $\sim 10^{43}$ erg s$^{-1}$,
this inner CLR  can well  be interpreted by
the photoionization model of Ferguson et al. (1997) whereas 
the outer part may have different physical conditions from that of
the NLR because the electron density in the outer region
($<10^2$ cm$^{-3}$: MTI98) is significantly lower than those in the NLR. 
Thus the outer CLR in this galaxy can be understood in terms of
the low-density CLR suggested by Korista \& Ferland (1989).

\section{DISCUSSION}

The arguments described in the previous section  suggest strongly that
there are three kinds of CLR; 1) the torus CLR ($r < 1$ pc),
2) the CLR associated with the NLR ($10 < r < 100$ pc), and 
3) the very extended CLR ($r \sim$ 1 kpc).
Therefore, if we take these three emission-line regions into account,
we may have a unified picture for the CLR of AGN.
Their  basic physical properties are summarized in Table 1.
A schematic illustration of the CLR is shown in Figure 3.
We mention that there is the large scatter in the
[FeVII]/[OIII] intensity ratio in both the S1s and the S2s.
This scatter suggests that the contribution of CLR
emission from the inner, extended, and very extended CLR may be
different from object to object.
Moreover, it should be remembered that a half of Seyfert nuclei
show no evidence for the CLR (Osterbrock 1977; Koski 1978). 
This may be attributed to a gas-rich
condition in the circumnuclear region, resulting in a lower ionization condition.
This diversity  may make it difficult to
construct a simple photoionization model for the CLR as well as
for the NLR itself (Ferland \& Osterbrock 1986).

In view of recent new observations and insights,
we discuss the nature of the three kinds of CLR in AGN.

1) The torus CLR: Given the current unified model, it is naturally considered that
clouds on the inner edges of dusty tori provide important sites
for the CLR (Pier \& Voit 1995).
A typical electron density is estimated to be $10^{7-8}$ cm$^{-3}$
(Pelat et al 1981; Ferguson et al. 1997; Pier \& Voit 1995). 
Since the emissivity of coronal lines is proportional to $n_{\rm e}^2$
under conditions of $n_{\rm e} < n_{\rm cr}$, 
the torus CLR can be the most luminous component.
This is indeed shown in Figure 1.
We also have to note that iron is often depleted in the interstellar medium.
However, since the inner edges of the tori are exposed to the intense
radiation field from the central engine and thus dust grains may be destroyed
(Pier \& Voit 1995). This is another reason why the torus CLR is more luminous
than that in the NLR. We also mention that the torus CLR consists of many small
ionized gas clumps though we illustrate it as shown in Figure 3. 

If we assume that the inner wall obeys a Keplerian rotation,
we obtain a typical line width 
FWHM $\simeq 2 v_{\rm rot} \simeq 1320 M_8^{1/2} r_{1}^{-1/2}$
km s$^{-1}$ where $M_8$
is the dynamical mass within a radius $r_1$ in units of $10^8 M_\odot$ and $r_1$
is the radius of the NLR in units of 1 pc (Pier \& Voit 1995).
This fiducial value is almost comparable with those of coronal lines
whenever they are broad 
(De Robertis \& Osterbrock 1984, 1986; Appenzellar \& \"Osreicher 1988;
Appenzellar \& Wagner 1991; Giannuzzo et al. 1995).
It has been known that some Seyfert nuclei and quasars have ionized regions
whose line widths are a few 1000 km s$^{-1}$ (Brotherton et al. 1994; Mason et al. 1996).
Since these line widths are intermediate between those of the NLR and the BLR,
it has been suspected that there is an intermediate line region (ILR)
in addition to the traditional NLR and BLR.
We propose that this ILR is just located at the inner wall of the tori
and thus the torus CLR may be associated spatially with the ILRs.

2) The CLR associated with the NLR (the clumpy CLR):
The recent {\it Hubble Space Telescope} observations of the NLR
in a number of nearby Seyfert nuclei have shown that the NLR consists of
a large number of gas clumps and thus the structure of the NLR turns out to be 
much more complex than what we thought (Wilson et al. 1993; Bower et al. 1995;
Capetti et al. 1996a, 1996b).
Therefore, we refer this CLR as the clumpy CLR.
It is naturally expected that  the cloud  surface facing to the
continuum radiation may be the major site of CLR.
MTI98 found that there is no correlation between [FeVII] and
an optical FeII feature at $\lambda$4570 which is presumed to arise from
warm neutral, or partially-ionized regions of gas clouds. 
Further, there is no correlation between
[FeVII] and [OI] (Murayama and Taniguchi, in preparation).
These properties imply that highly ionized gas clumps are decoupled from low-ionization
gas clumps. We may interpret that the CLR arises from matter-bounded ionized clumps
while low-ionization lines arises mostly from ionization-bounded gas clumps\footnote{
Recently, measuring the electron temperatures of both O$^{2+}$ and N$^{+}$
regions in the extended emission-line regions of several 
Seyfert galaxies, Wilson et al. (1997) proposed that the [OIII] emission
arises mostly from the matter-bounded clouds while the [NII] emission arises
from ionization-bounded clouds because the observed electron temperature difference
between the [OIII] and [NII] regions
is too large (e.g., $\sim$ 5000 K) to be accounted for in terms of
photoionization of ionization-bounded clouds. On the analogy of this finding,
we consider that the NLR also consists of the two kinds of clumps.}.
Since the [FeVII] emitting region should be 
exposed to the harder and stronger radiation field than the [OIII] region,
the radiation pressure exerted from the central engine is higher 
for high-ionization gas clumps than for the low-ionization ones 
and thus the high-ionization clumps
may  be accelerated more efficiently, leading to the systematic
blueshift of the high-ionization clumps with respect to the lower ones.
This property has been often observed in many AGN
(Grandi 1978; Appenzellar \& \"Osreicher 1988; Gaskell 1982; Wilkes 1984).

3) The extended CLR:
 The very extended CLR ($r \sim$ 1 kpc) is found in  both
NGC 3516 (Golev et al. 1994) and Tololo 0109$-$383 (MTI980.
If the interstellar medium consists of very low-density gas clouds (e.g.,
$n_{\rm H} \sim 1$ cm$^{-3}$), the high ionization condition can be 
achieved (Korista \& Ferland 1989). Therefore, it is reasonable that
some Seyfert nuclei have such the very extended CLR.
The extended CLR seems to be related to the so-called
extended ($r \sim$ 1 - 10 kpc) narrow  line region (ENLR; Unger et al. 1987) which is 
thought to be interstellar medium photoionized by the continuum radiation
from the central engine.
The CLR may be lower-density parts of the ENLR because the CLR needs higher
ionization condition than the typical [OIII] emitting region.

\vspace {0.5cm}

We would like to thank B. Vila-Vilalo for useful discussion.
TM was supported by the Grant-in-Aid for JSPS Fellows
by the Ministry of Education, Science, Sports, and Culture.
This work was financially supported in part by Grant-in-Aids for the Scientific
Research (No.\ 0704405) of the Japanese Ministry of
Education, Science, Sports, and Culture.

\clearpage

\begin{deluxetable}{lcccc}
\tablecaption{The three-component model for the CLR}
\tablehead{
 & \colhead{$r$ (pc)} & \colhead{$n_{\rm H}$ (cm$^{-3}$)} & \colhead{FWHM (km s$^{-1}$)} 
& \colhead{Associated}  \nl
 &  & &  & \colhead{Emission-line region}  \nl
}
\startdata
Torus CLR & $\sim$ 1 & $\sim 10^7$ - $10^8$ & 1300\tablenotemark{a} & ILR\tablenotemark{b} \nl
Clumpy CLR & $\sim$ 1 - 100  & $\sim 10^3$ - $10^6$ & 400 - 750 & NLR\tablenotemark{c} \nl
Extended CLR & $\sim$ 1000 & $\sim 1$ & $<$ 50\tablenotemark{d} & ENLR\tablenotemark{e} \nl
\enddata
\tablenotetext{a}{An observed FWHM is 2$v_{\rm rot}$sin$i$ where $i$ is the angle 
between the line of sight and the rotational axis of the torus. 
When we observe the torus from
a face-on view, the FWHM should be a virial line width, 
660 $M_8^{1/2} r_1^{-1/2}$ km s$^{-1}$.
Such narrow line widths of the CLR are observed in some Seyfert nuclei
(De Robertis \& Osterbrock 1984, 1986; Giannuzzo et al. 1995).}
\tablenotetext{b}{Intermediate line region.}
\tablenotetext{c}{Narrow line region.}
\tablenotetext{d}{Unger et al. (1987)}
\tablenotetext{e}{Extended narrow line regions. Another term, extended emission-line regions (EELRs)
is also used.}
\end{deluxetable}
\clearpage


\clearpage


\figcaption{%
Frequency distributions of the [FeVII]$\lambda$6087/[OIII]$\lambda$5007
intensity ratio between the S1s and the S2s.
}

\figcaption{%
Frequency distributions of the [OIII]$\lambda$5007 luminosity
between the S1s and the S2s.
}

\figcaption{%
A schematic illustration of the three-component model for the CLR.
Note that the torus CLR consists of many small ionized gas clumps
like the clumpy CLR in the NLR.
}

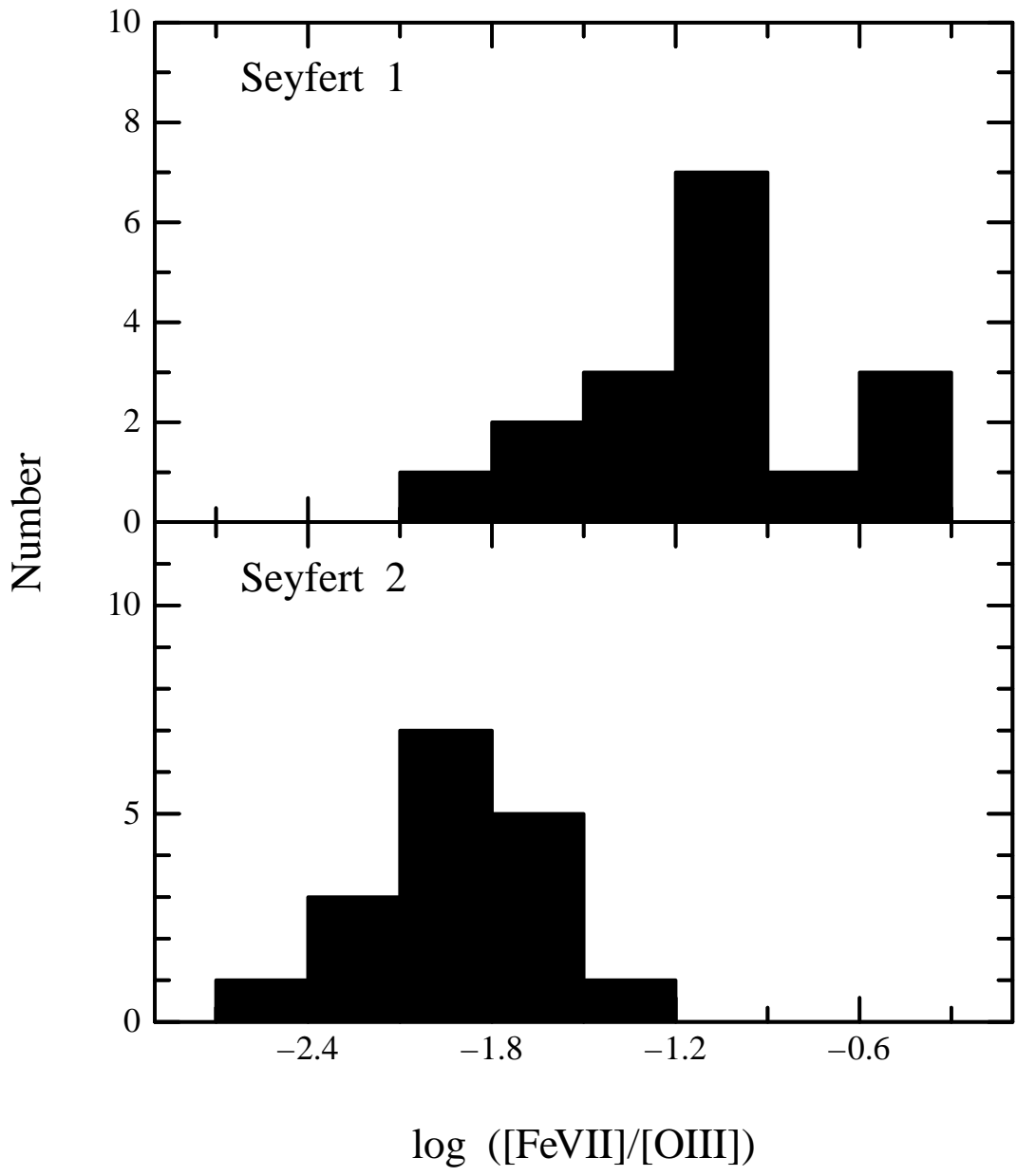
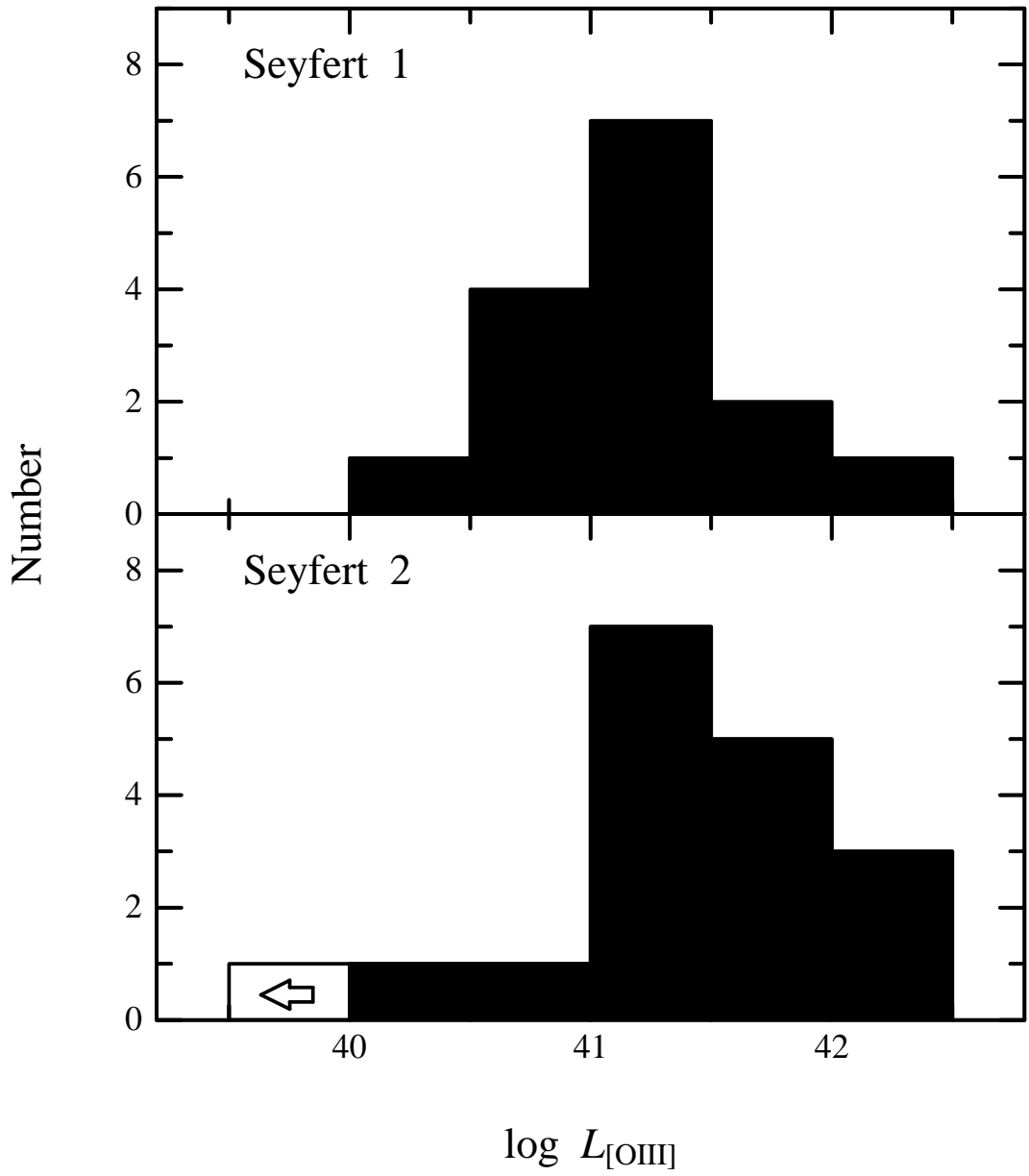
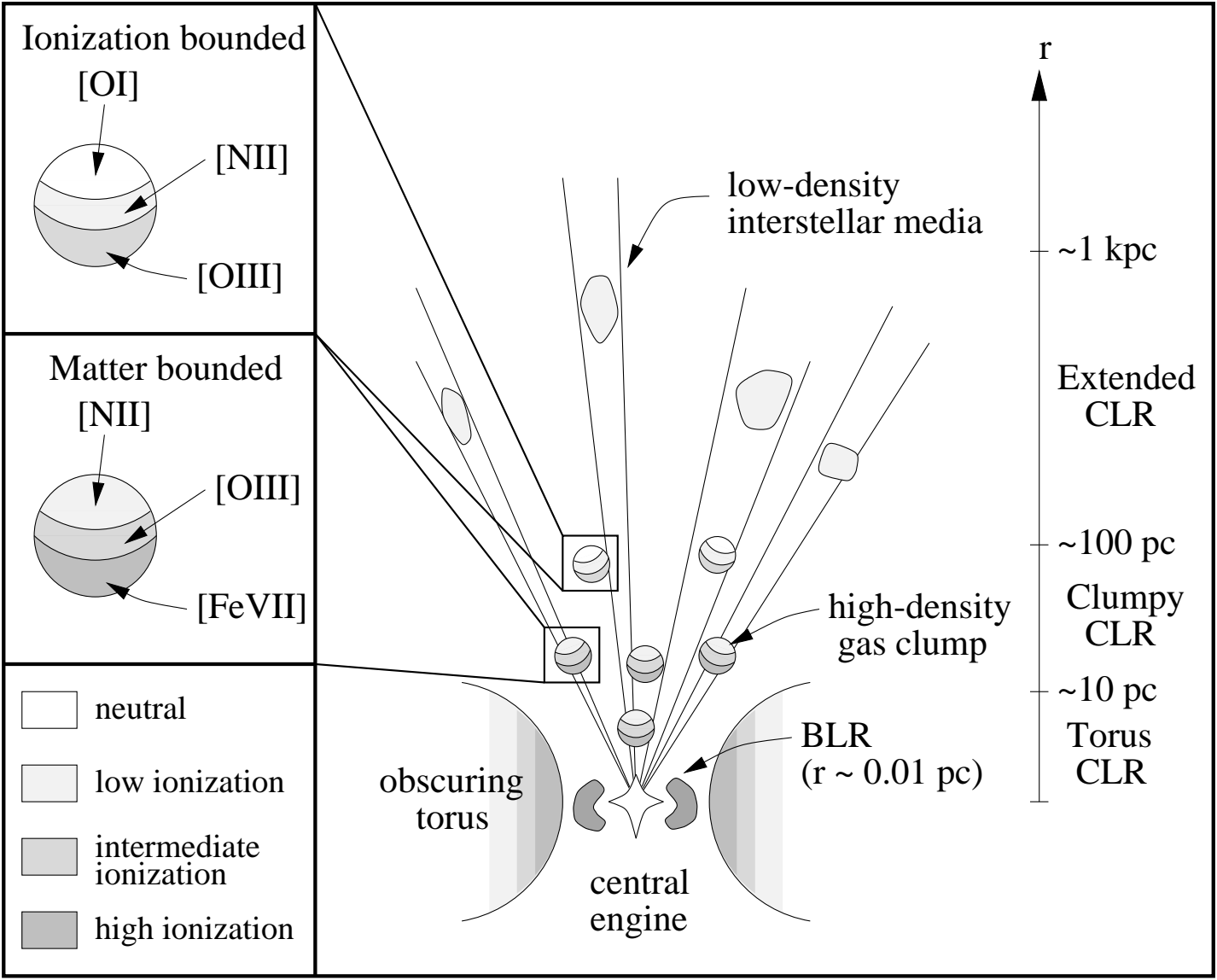
\end{document}